\documentclass[showpacs,showkeys,preprintnumbers,superscriptaddress,amsmath,amssymb,latexsym,twocolumn]{revtex4}
\usepackage{graphicx}
\usepackage[latin1]{inputenc}
\usepackage[english]{babel}

\def\beq{\begin{equation}}
\def\eeq{\end{equation}}

\begin{document}

\title{Excised acoustic black holes: the scattering problem in the time domain}

\author{C. Cherubini}
\affiliation{Facolt\`a di Ingegneria, Universit\`a  Campus Biomedico di Roma,
Via Longoni 83, 00155 Roma, Italy and \\
I.C.R.A., Universit\`a di Roma \lq\lq La Sapienza\rq\rq, 00185 Roma, Italy}
\author{F. Federici}
\email{fr.federici@sns.it}
\affiliation{NEST-INFM and Classe di Scienze, Scuola Normale
  Superiore, Piazza dei Cavalieri 7,  56126 Pisa, Italy}
\author{S. Succi}
\affiliation{Istituto per le Applicazioni del Calcolo, CNR, Viale del
  Policlinico 137, 00161 Roma, Italy}
\author{M. P. Tosi}
\affiliation{NEST-INFM and Classe di Scienze, Scuola Normale
  Superiore, Piazza dei Cavalieri 7,  56126 Pisa, Italy}


\begin{abstract}

The scattering process of a dynamic perturbation impinging on a draining-tub model of an acoustic black hole is numerically solved in the time domain. Analogies with real black holes of General Relativity are explored by using recently developed mathematical tools involving finite elements methods, excision techniques, and constrained evolution schemes for strongly hyperbolic systems. In particular it is shown that superradiant scattering of a quasi-monochromatic wavepacket can produce strong amplification of the signal, offering the possibility of a significant extraction of rotational energy at suitable values of the angular frequency of the vortex and of the central frequency of the wavepacket. The results show that theoretical tools recently developed for gravitational waves can be brought to fruition in the study of other problems in which strong anisotropies are present.

\end{abstract}


\maketitle

\section{Introduction}

General Relativity (GR) deals with dynamical deformations of space and
time, making a massive use of tensor calculus in parallel with continuum mechanics. In recent years the exploration of Einstein's theory has received a strong acceleration from the growing efforts in high-precision experiments involving gravity. Ground-based and space-born interferometers like LIGO, VIRGO, TAMA, GEO600, and LISA should at long last allow detection of gravitational waves. The analysis of the expected signals will require precise theoretical templates to be compared with the measured waveforms via pattern matching. For this reason a variety of theoretically simulated scenarios are needed, as obtained by numerical integration of the systems of partial differential equations (PDE's) of GR. The striking developments of Numerical Relativity over the last ten years, which rely on presently available large-scale computational resources, have shown that some engineering tools such as the finite-elements techniques can be fruitfully adopted to these ends. On the other hand, the need to deal with event horizons, Cauchy horizons, and infinite-curvature singularities during the numerical evolution still poses serious theoretical and numerical challenges.
	
In the past a very simple and yet profound
analogy between GR and Newtonian physics has been noticed
\cite{ref41, Jacobson, Visser} and it can be hoped that the former may benefit from the wide body of knowledge which is available for the latter. In particular, given a perfect barotropic and irrotational Newtonian fluid, perturbations of the velocity potential with respect to a background solution have been analysed and shown to satisfy a second-order linear hyperbolic equation with non-constant coefficients. By algebraic manipulations this wave equation can be rewritten as a Klein-Gordon field 
propagating on a pseudo-Euclidean four-dimensional Riemannian manifold: an induced effective gravity is at work in the fluid, in which the speed of sound plays the role of the speed of light.

GR, however, prescribes the existence of black holes. Surprisingly, some of the induced geometries in the analog formulation have the same structure as those of a black hole. Sound waves crossing the horizon of these acoustic counterparts will never come back out. At this level a practical problem arises. A black hole acts on light propagation as an anisotropic medium having an infinite refractive index at the horizon, and from studies of strongly anisotropic media, one can anticipate that a numerical study of this problem will face serious difficulties. One may also ask whether black-hole behavior could emerge in other fields where hyperbolic wave equations are met, for instance in dealing with electromagnetism of anisotropic media or with perturbations of elastic objects.

Fortunately, Numerical Relativity has developed the proper
instruments to deal with this type of problems, through the
	use of excision techniques together with a constrained
	evolution scheme for strongly hyperbolic problems. The main
	purpose of the present work is to use this methodology in the
	study of the simplest acoustic black hole, {\it i.e.} the
	draining bathtub model of Visser \cite{Visser}. While much can be borrowed from the extensive literature in Numerical Relativity, the task of developing the apparatus needed to pursue this black-hole analog on fully quantitative grounds by numerical simulations does not seem to have been previously addressed. The relevance of these studies to superradiant extraction of rotational energy from a vortex in a Bose-Einstein condensate has been discussed elsewhere \cite{FCST}.

	The plan of the paper is as follows. In Sect. II we present a detailed analytical study of the draining bathtub geometry in analogy with standard treatments of a rotating Kerr black hole. In Sect. III we recast the hyperbolic equation to a constrained symmetric system of first-order PDE's and solve a numerical scattering problem on the basis of a finite-element method. In Sect. IV we analyse the time evolution of both compact and quasi-monochromatic sound-wave pulses, report on tests of the validity of the proposed procedure, and pay special attention to the scattering of a quasi-monochromatic wavepacket in the superradiant regime. Finally, Sect. V concludes the paper with a discussion of the main results and an illustration of future perspectives regarding vortices in superfluids.

\section{The model}

We aim at investigating sound-wave scattering processes from a sonic
black hole by studying the time evolution of a scalar field in the
presence of a vortex in the fluid flow. It is well known
\cite{ref41,Visser} that
sound propagation in such a fluid is described by a wave equation
which, under the assumption of long-wavelength perturbations, can be
mapped onto a Klein-Gordon equation associated with an effective
relativistic curved space-time background (an acoustic metric). We
focus our attention on the acoustic metric associated to the
draining-bathtub model introduced by Visser \cite{Visser} for a rotating
acoustic black hole.

The model is based on irrotational barotropic and incompressible
Euler's equations describing a (2 + 1)-dimensional flow with
a sink at the origin. The flow is taken as vorticity-free
(apart from a possible $\delta$-function contribution at the vortex
core) and angular momentum is conserved. These constraints
imply that the density of the fluid, the background pressure,
and the speed of sound $c$ are constant throughout the flow. The
background velocity potential must therefore have the form
\begin{equation}
\psi(r,\phi)=-c a \log(r/ a)+\Omega a^2\phi,
\label{POTENT}
\end{equation}
where $a$ is a length scale associated with the \lq\lq radius\rq\rq" of the vortex horizon and $\Omega$ is the (constant) rotation frequency.
	
Such a velocity potential, being discontinuous on going through $2\pi$ radians, is a multivalued function. Therefore it must be interpreted as being defined patch-wise on overlapping regions around the vortex core at $r = 0$. The velocity of the fluid is then given by
\begin{equation}
\vec v(r,\phi)=\nabla \psi(r,\phi)=\frac{-c a \hat r+\Omega a^2\hat\phi}{r}\,.
\label{VEL}
\end{equation}
The acoustic metric associated to this configuration is
\begin{eqnarray}
\label{METRIC1}
{\rm d}s^2=&-&\left(c^2-\frac{ a^2 c^2+ a^4\Omega^2}{r^2}\right){\rm d}t^2+
\frac{2c a }{r}{\rm d}t\,{\rm d}r\\
&-&2\Omega a^2 {\rm d}t\, {\rm
  d}\phi+{\rm d}r^2+r^2{\rm d}\phi^2+{\rm d}z^2\nonumber.
\end{eqnarray}
It is easily shown that the metric (\ref{METRIC1}) exhibits an acoustic
 event horizon located at $r = a$, where the radial component of the
 fluid velocity exceeds the speed of sound, and an ergosphere at 
$r=r_{\rm erg}=a\sqrt{1+\Omega^2a^2/c^2}$ .
 A detailed analysis is given in Appendix \ref{appA}.

Linear perturbations $\psi^{(1)}\equiv
\Psi$ of the velocity potential satisfy the massless Klein-Gordon scalar wave equation
$\nabla^\mu\nabla_\mu\Psi=0$ on this background, {\it i.e.}
\begin{eqnarray}
\label{SCALEQ}
&&\left[-\frac{1}{c^2}\frac{\partial^2}{\partial t^2}+\frac{2 a}{cr}\frac{\partial^2}{\partial t\partial r}-\frac{2 a^2\Omega}{c^2 r^2}\frac{\partial^2}{\partial t \partial \phi}+\left(1-\frac{ a^2}{r^2}\right)\frac{\partial^2}{\partial r^2}+\right.\nonumber\\
&&\left.+\frac{2 a^3\Omega}{cr^3}\frac{\partial^2}{\partial r\partial \phi}+
{\frac {{c}^{2}{r}^{2}-{
      a}^{4}{\Omega}^{2}}{{c}^{2}{r}^{4}}}\frac{\partial^2}{\partial
  \phi^2}+\frac{\partial^2}{\partial z^2}\right.\\
&&\left. +{\frac {r^2+{ a}^{2}}{{r}^{3}}}\frac{\partial}{\partial r}-
\frac{2 a^3\Omega}{cr^4}\frac{\partial}{\partial \phi}
\right]\Psi=0.\nonumber
\end{eqnarray}
This PDE is strongly hyperbolic, and its numerical integration is best performed by resorting to the tools of Differential Geometry (and thus of General Relativity).

Equation (\ref{SCALEQ}) is completely separable and we analyse it in the 
frequency domain by setting
$\Psi=e^{-i\omega t}e^{im\phi }e^{ikz} P(r)$, where $k$ and the
integer $m$ are the axial and azimuthal wavenumbers, respectively. 
The result is a  second-order ordinary differential equation (ODE):
\begin{eqnarray}
&&\left(1-\frac{a^2}{r^2}\right)
\frac{{\rm d}^2 P}{{\rm d}r^2}
+\left(\frac{ca^2+2ia^3\Omega m}{cr^3}+
\frac{c-2ia\omega}{cr}\right)
\frac{{\rm d}P}{{\rm d}r}+\nonumber\\
&&\left(\frac{\omega^2}{c^2}-k^2-\frac{m^2}{r^2}
-\frac{2i a^3\Omega m}{cr^4}\right.\\
&&\left.-\frac{2a^2\Omega m\omega}{c^2 r^2}+
\frac{a^4\Omega^2m^2}{c^2r^4}\right)P=0.\nonumber
\label{RADIALEQ}
\end{eqnarray}
Here $m$ is taken as integer in order to avoid 
polydromy problems, while $k$ is a 
real number fixed by the boundary conditions along the $z$ axis.
It may be interesting to notice that by the Frobenius method for
 ODE's it can be shown that the indicial equation 
computed at  $r= a$ has two roots, $\lambda_1=0$ and 
$\lambda_2=\frac{i a}{c}\left(\omega-m\Omega\right)$.
The latter expression leads to  the well known 
condition for rotational superradiance: 
$0 <\omega < m \Omega$ for $\omega\in{\cal R}$, as shown by the
asymptotic study of the ODE for $r=a$ and $r\rightarrow\infty$ 
\cite{Wald,Basak1,Berti}. 

Time dependent modes with $m=0$ can be  analytically 
solved in terms of hypergeometric functions. In this special case
Eq. (\ref{RADIALEQ})  can be cast in a 
 Schroedinger-like form, with a 
\lq\lq dressing\rq\rq potential outside the horizon. 
Following a standard procedure 
\cite{TEUKOLSKY}, by rescaling the field as
\begin{eqnarray}
P(r)=S(r)\exp\left\{{-\frac12\int\left[ \frac{1}{r}-\frac{2i\omega a r}
{c(r^2- a^2)}\right]{\rm d}r}\right\}
\label{pr}
\end{eqnarray} 
and by introducing the \lq\lq tortoise-like \rq\rq coordinate 
\begin{equation}
r_*=\int\frac{r^2}{r^2- a^2}{\rm d}r
\label{rtort}
\end{equation}
which maps the radial coordinate from $[ a,+\infty]$ to $[-\infty,+\infty]$,
we obtain the equation
\begin{equation}
\frac{\rm d^2 S}{\rm d r_*^2}+\left[\frac{\omega^2}{c^2}-V(r)
\right]S=0
\label{ss}
\end{equation}
where
\begin{equation}
\label{pot}
V(r)={k}^{2}-\frac54\,{\frac {{ a}^{4}}{{r}^{6}}}-\frac14\,\left({\frac {1+4\,{k}^{2}{
 a}^{2}}{{r}^{2}}}\right)+\frac32\,{\frac {{ a}^{2}}{{r}^{4}}}.
\end{equation}
It should be noted that $r_*$ is a monotonic function of $r$ and
differs appreciably  from 
$r$ only in the vicinity of the horizon.

\begin{figure}
\centering
\includegraphics[width=8cm]{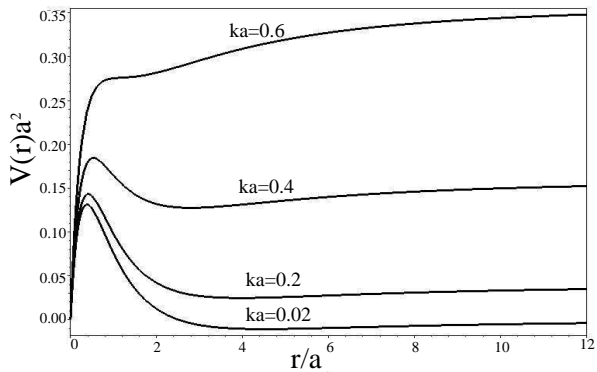}
\caption{The horizon-dressing potential $V(r)a^2$ is plotted as a
  function of $r/a$ for $ka = 0.02$, 0.2, 0.4, and 0.6 (from bottom to top).}
\label{fig0}
\end{figure}

In Fig. \ref{fig0} the potential $V(r)$ is plotted as a function of $r$
for several values of $k$. 
Waves of a given frequency $\omega$ are scattered
by the real potential $V(r)$ and can only partially tunnel 
towards the horizon for  $k\lesssim 0.5$. 
As $k$ increases further, the potential becomes more and more
 attractive and the back-reflecting barrier disappears.

In the general case $m\neq0$ the equation is complex, 
and a simple Schroedinger-like picture no longer applies.
Several  numerical studies 
of the wave equation (\ref{SCALEQ}) 
have recently been performed in the frequency domain for $k=0$ \cite{Basak1,Berti}. 
However, these studies are based on a radial equation which differs
from Eq. (\ref{RADIALEQ}).
New time and angle coordinates have been introduced in order to
minimize the dragging effects and to work in the region outside the
horizon only. The physical interpretation of the results is not
straightforward as the azimuthal quantum number associated with the
new angular coordinate reflects different geometrical properties from
those of the cylindrical angle $\phi$. 
This is a well-known problem in Kerr black hole physics \cite{Laguna}.  
In order to avoid  complications with the angular
coordinate, we shall keep the original cylindrical coordinate
system involving horizon-penetrating coordinates. A perturbative
formalism using these coordinates is 
highly appropriate in connection with numerical relativity simulations
using black-hole excision.

\section{Theoretical formulation of the scattering process}

We solve Eq. (\ref{SCALEQ}) numerically in the time domain by
 resorting to some of the mathematical 
prescriptions that have been developed by Scheel {\it et al.} 
\cite{TEUK0} for the  
numerical integration of massless scalar-field 
perturbations on a rotating Kerr black hole background.

The starting point is to introduce conjugate 
quantities of the scalar field $\Psi$:
\begin{eqnarray}
\Phi_i=\frac{\partial\Psi}{\partial x^i},\,\quad
\Pi=-\frac{1}{\alpha}\left(\frac{\partial\Psi}{\partial
  t}-\beta^i\Phi_i\right),
\label{15}
\end{eqnarray}
where $\beta^{i}=\left(acr^{-1},-a^2\Omega,0\right)$ and $\alpha=c$ 
(see Appendix \ref{appA}).
These are used in  Eq. (\ref{SCALEQ}) to obtain a symmetric
first-order hyperbolic system. We next 
factorize the dependence on $z$ and $\phi$ by setting 
$\Psi=\psi_1(t,r)e^{im\phi}e^{ikz}$, 
$\Pi=\pi_1(t,r)e^{im\phi}e^{ikz}$, 
$\Phi_1=\phi_1(t,r)e^{im\phi}e^{ikz}$,
$\Phi_2=im\Psi$, and $\Phi_3=ik\Psi$. 
These transformations allow us to reduce the computational 
demands. The wave equation (\ref{SCALEQ}) is then replaced by a 
first-order set of coupled PDE's,
\begin{eqnarray} 
\label{pdeset}
&&\partial_{t}\pi_1+c\partial_{r}\left(\xi_{1}- a\pi_{1}/r\right)=
\left(ac-im a^2\Omega\right)\pi_1/r^2+\nonumber\\
&&+c\left(k^2+m^2/r^2\right)\psi_{1} -c\xi_{1}/r \nonumber\\
&&\partial_{t}\psi_{1} -c\partial_{r}\left(a\psi_{1}/r\right)=
\left(ac-im a^2\Omega\right)\psi_{1}/r^2-c\pi_{1} \\
&&\partial_{t}\xi_{1}+c\partial_{r}\left(\pi_{1}-a\xi_{1}/r\right)=
2im a^2 \Omega\psi_{1}/r^3-ima^2\Omega\xi_{1}/r^2\nonumber\, 
\end{eqnarray} 
associated with the constraint 
\begin{equation}
\vert C\vert=\vert\partial_r\psi_1-\phi_1\vert=0
\end{equation}
following from the definition of $\phi_1$. 
Deviations of this quantity from zero provide a useful indicator
of the quality of the numerical results.

Following the standard prescription 
for scattering processes in Kerr black holes \cite{Laguna}, 
we choose as initial data a Gaussian pulse centered at $r=r_0$ and 
 modulated by a monochromatic wave, 
\begin{equation}
\psi_1(0,r)=A\exp[{-{(r-r_0+ct)^2}/{b^2}-i\sigma(r-r_0+ct)/c}]\vert_{t=0}
\end{equation}
together with the associated initial conditions 
for $\pi_1$ and $\phi_1$.
The corresponding power spectrum is 
$P(\omega)=P_{{\rm max}}\exp{\left[-(\omega-\sigma)^2b^2/4c^2\right]}$. 
The modulated pulse thus has a Gaussian frequency distribution
centered at $\omega=\sigma$ and the choice of $\sigma$ in the 
range $0<\sigma<m\Omega$ with $m\geqslant 1$
corresponds to the superradiant regime.

It is well known that the numerical integration of wave equations in
cylindrical geometry must handle reflected
waves close to the inner boundary (see {\it e.g.} Koyama \cite{Koyama}).
Similar difficulties arise in dealing with black holes,
 where theoretically
correct boundary conditions are frequently in need of {\it ad hoc}
modifications such as \lq\lq background subtraction\rq\rq 
and \lq\lq peeling off properties of the field 
($1/r^{n}$ fall-offs)\rq\rq at large distance
 \cite{Thornburg}. The
best way to solve the problem is to use penetrating coordinates or,
better yet, the Kerr ingoing coordinates in Eq. (\ref{COORDYS}).
These permit to extend the field integration inside the horizon via
the so-called excision technique. Inside the horizon, the light cones
point towards the singularity, so that the interior region is causally
disconnected from the region outside the horizon. Thanks to this
property, one can impose a generic well-behaved boundary condition
inside the horizon and sidestep the problem of constraint violations
inside the acoustic black hole. This procedure is computationally very
demanding, as it relies upon a very fine discretization around the
sonic horizon. Owing to requirements of high resolution, long
integration times, and distant outer boundary, even the simple
Klein-Gordon equation in Kerr black-hole geometry has only recently
been successfully implemented with a powerful multi-processor code
\cite{TEUK0}. Following Scheel {\it et al.} \cite{TEUK0}, we have implemented the excision technique for an acoustic black hole and chosen the boundary conditions as follows.

We have set no boundary conditions on the inner boundary $r_{\rm in}$ , with
$0<r_{\rm in}<a$. Regarding the outer boundary $r_{\rm out}$, 
the proper criterion is to adopt conditions which will not violate constraints, will not spontaneously generate unphysical waves, and will not significantly back-reflect any incoming signal. In our case we have first defined the directional derivatives of $\Psi$ along the principal null directions given in Eq. (\ref{PNDS}), namely
\begin{eqnarray}
u^{+}&=&n^\mu\partial_\mu \Psi\propto\left(\pi_1+\phi_1\right)
e^{im\phi}e^{ikz},\nonumber\\
u^{-}&=&l^\mu\partial_\mu \Psi\propto \left(\pi_1-\phi_1\right)e^{im\phi}e^{ikz}\,\, \,.
\end{eqnarray}
At large distances we must impose a purely outgoing boundary 
condition, in order to avoid backscattered radiation
which would contaminate the signal received at an observational point
located closer to the vortex: consequently we set $u^-=0$,
which leads to no boundary conditions for $\psi_1$ and to
$\phi_1=\pi_1$. In the limit of a flat space-time ($a=0$),
 from Eq. (\ref{15}) this condition becomes
$c\partial_r\psi_1+\partial_t\psi_1=0$ and yields $\psi_1=h_2(r-ct)$,
which is the familiar condition for zero ingoing (left-moving)
waves. These boundary conditions did not lead to constraint violations nor generation of spurious signals or of significant reflected waves during the numerical integration of the field equations. 
Moreover, by stopping the integration before any physical signal has 
causally reached the outer boundary, even tiny reflected waves are 
completely eliminated.

\section{Numerical results}

 The integration of the field equations is performed by using  finite
elements techniques instead of the standard finite-difference ones.
We have used in particular the nonlinear engine of 
FEMLAB$^{\copyright}$, adopting a fifth-order polynomial Lagrange
element with a uniform mesh interval with $\Delta r= 0.05$. 
Although this mesh is
not very fine, the use of higher-order elements yields very good
results. The integration is performed for $r\in[0, 150a]$ while time
is in the range $t\in[0, 150a/c]$ with an optimal time step chosen by
the nonlinear solver. In order to avoid constraint violations and to
secure stability, we have used finite elements of lower (third
or even second) order close to the outer boundary. In our numerical
analysis we have mainly focused on the scattering of a 
\lq\lq quasi-monochromatic\rq\rq (large $b$) wavepacket both inside
and outside the superradiant regime, although the case of a spatially
\lq\lq quasi-localised\rq\rq (compact) wavepacket has also been
examined.

In the calculations we have adopted values of the parameters that can
be related to realistic superfluid systems \cite{FCST} 
and examined both the
non-superradiant and the superradiant regime by suitable choices of
the value of $\sigma/\Omega$. The angular frequency $\Omega$
 was analysed in the range
$0.14 < \Omega a/c < 14$, and the values $m = 0$ and $1$, $ka =
 0.02$, and $A = 0.3c$ were selected
together with $r_0=10a$ and $b=0.5a$ for the
quasi-localised packet and $r_0 = 50a$ and $b = 10a$ 
for the quasi-monochromatic one. 
Code units corresponding
to $a = 1$ and $c = 1$ have been used.

\begin{figure}
\centering
\includegraphics[width=9cm]{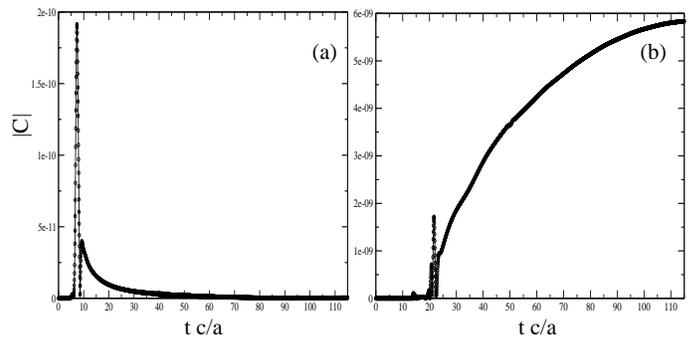}
\caption{Constraint violations at the inner (a) and outer (b) boundary
   for $m=0$ in the case of a compact wavepacket ($b=0.5a$,
   $\sigma=0$) for $\Omega=1.4 a/c$ and $r_0=10a$.}
\label{Fig1}
\end{figure}

\begin{figure}
\centering
\includegraphics[width=9cm]{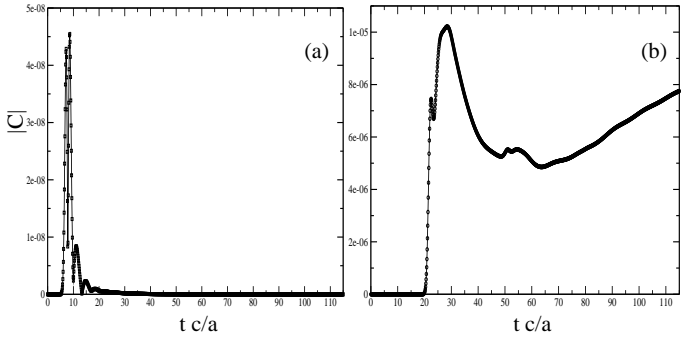}
\caption{Constraint violations at the inner (a) and outer (b) boundary
  for $m=1$. The parameters used in the calculation are as 
in Fig. \ref{Fig1}.}
\label{Fig2}
\end{figure}

As an example of our diagnostics, Fig.\ref{Fig1} and Fig.\ref{Fig2} plot the evolution 
of $\vert C\vert$  as a function of $t$ for $m=0$ and $m=1$ at the inner  and  outer
boundaries. These figures show that our boundary conditions
preserve the constriants to a very satisfactory degree. 
In both cases the perturbation does not grow indefinitely in time:
the acoustic black hole appears to be stable against these
perturbations. The
evolution of $\vert C\vert$ for $m=1$ at the outer boundary confirms
that  more attention has to be paid
 to possible
constraint violations in this case (as already found in
\cite{TEUK0}). However, 
such violations in Fig. \ref{Fig2}(b) are still negligible.
In both cases, constraint violations remain within
the range $10^{-5}$ to $10^{-10}$, which means
that causality violations, if present, are practically irrelevant.

\begin{figure}
\centering
\includegraphics[width=8cm]{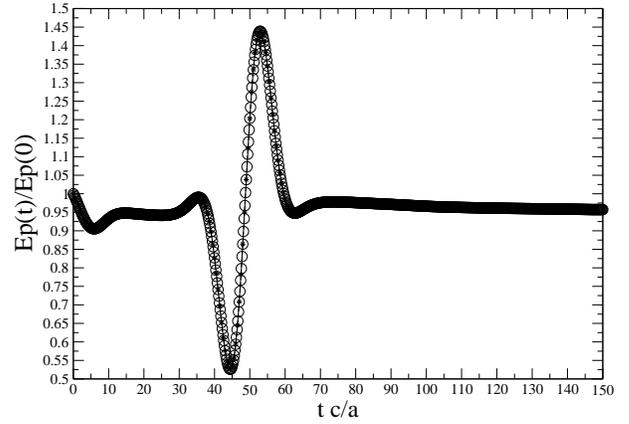}
\caption{The energy $E_{\rm p}(t)$ as a function of $tc/a$, normalised to its
initial value $E_{\rm p}(0)$, as calculated with two different choices
of the mesh (see text). The parameters used in the calculation
are $\sigma=0.07c/a$, $\Omega=0.14c/a$, $b=10a$, and $r_0=50a$.}
\label{Fig3}
\end{figure}

As a further illustration of the stability of our results, we compare
in Fig. \ref{Fig3} the values for the time-dependent energy of the
wavepacket calculated with the standard mesh of $0.05$ with those
obtained on the finer mesh with $\Delta r= 0.003$. 
There evidently is complete agreement
between the two sets of results, even in the time range of the
scattering process where the energy of the wavepacket is varying very
rapidly in time.

\begin{figure}
\centering
\includegraphics[width=8cm]{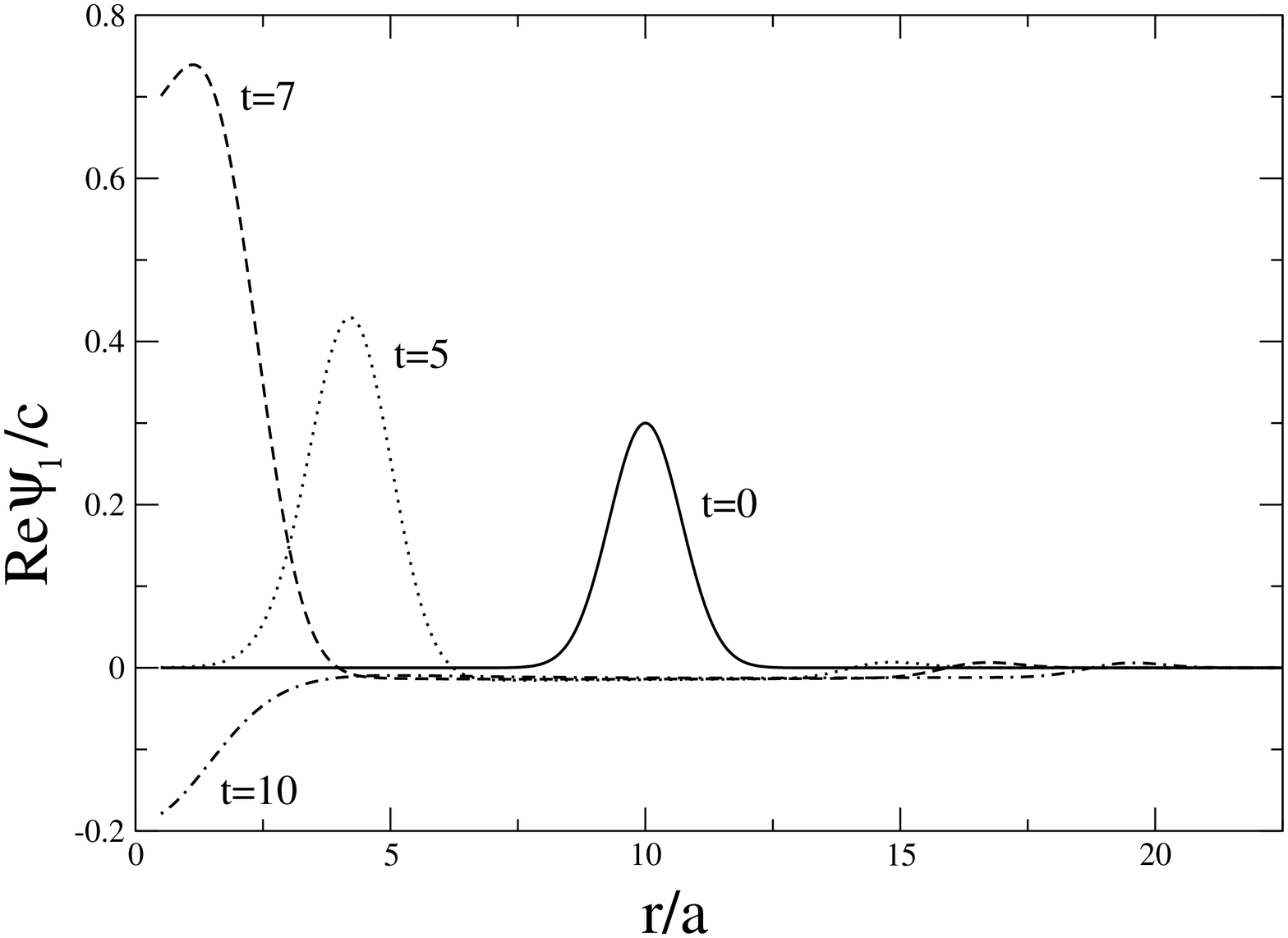}
\caption{Snapshots of the time evolution of the perturbation $\psi_1$
  representing a compact wavepacket 
  at $t=0$ (solid line), $t=5$ (dotted line), $t=7$ (dashed
  line), and $t=10$ (dot-dashed line), as a function of the
  distance $r$ from the the vortex center for the case $m=0$ ($t$ in
  units of $a/c$). 
 The solid line shows the initial
  Gaussian pulse travelling towards the vortex. The parameters used 
in the calculation are as 
in Fig. \ref{Fig1}. }
\label{Fig4}
\end{figure}

\begin{figure}
\centering
\includegraphics[width=8cm]{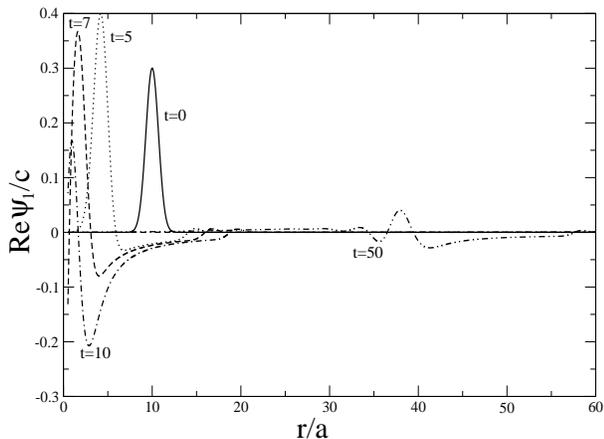}
\caption{Snapshots of the time evolution of the perturbation $\psi_1$
representing a compact wavepacket 
  at $t=0$ (solid line), $t=5$ (dotted line), $t=7$ (dashed
  line), and $t=10$ (dot-dashed line), as a function of the
  distance $r$ from the the vortex center for the case $m=1$ ($t$ in
  units of $a/c$). The parameters used in the calculation are as 
in Fig. \ref{Fig1}. }
\label{Fig5}
\end{figure}

We can now describe the main physical results of the numerical integration. Fig.\ref{Fig4} and Fig.\ref{Fig5} report snapshots of the time evolution
of the real part of $\psi_1$ as a function of $r/a$ for $tc/a=0$, $5$,
$7$ and $10$ in
the case $m=0$ and $m=1$, respectively. The peak in the signal at
$t=0$ shows the ingoing Gaussian pulse.
At $t=5a/c$  the pulse starts being affected by the dressing potential and
 the radiation is backscattered near $t=7$ for $m=0$ and $t=10$ for $m=1$.
The small reflected wave then relaxes towards the steady state 
(${\rm Re}\Psi_1 =0$) after the 
characteristic ringing-mode oscillations.

\begin{figure}
\centering
\includegraphics[width=10.5cm]{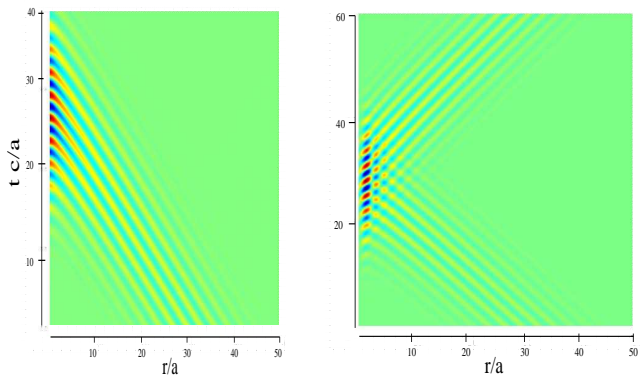}
\caption{A density plot of the real part of $\psi_1$ 
in the r-t plane for $m=0$ (left) and $m=1$ (right) in the case of a
quasi-monochromatic wavepacket. The calculations refer to $\Omega=3
c/a$ and $\sigma=0.7\Omega$ (color online).}
\label{Fig6}
\end{figure}

Figure \ref{Fig6} shows a 
density plot of the real part of $\psi_1$ 
in the range $r/a\in[0,60]$ and $tc/a\in[0,60]$ for the case $m=0$ (no
superradiance, left) and $m=1$ (superradiance, right).

The initial Gaussian pulse moves towards the vortex horizon placed at $r=a$, its trajectory
being bent by the potential outside the horizon. The bending of the
pulse trajectory, with light cones heading towards
the horizon, is consistent with similar findings in numerical
relativity \cite{calabrese}. Qualitatively similar results are
obtained in the non-superradiant regime for $m=1$ by choosing
$\sigma=1.2 \Omega$.

\begin{figure}
\centering
\includegraphics[width=8cm]{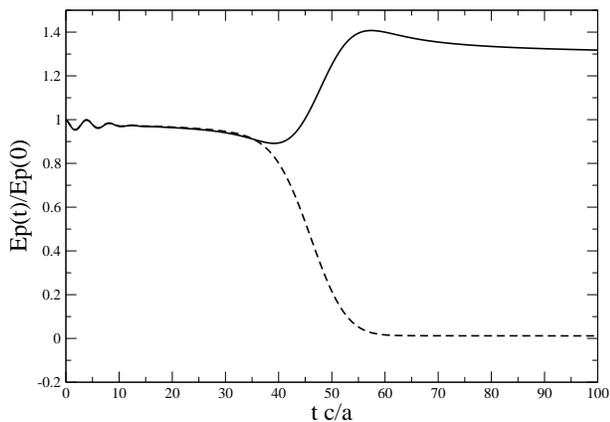}
\caption{Time evolution of the energy gain $E_{\rm p}(t)$ of the
  wavepacket (in units of its initial value $E_{\rm p}(0)$) for
  $\sigma=0.7c/a$ and $\Omega=1.4c/a$, for the superradiant case ($m=1$,
  solid line) and the non-superradiant case ($m=0$, dashed line). The
  time $t$ is in units of $a/c$.}
\label{Fig7}
\end{figure}

\begin{figure}
\centering
\includegraphics[width=8cm]{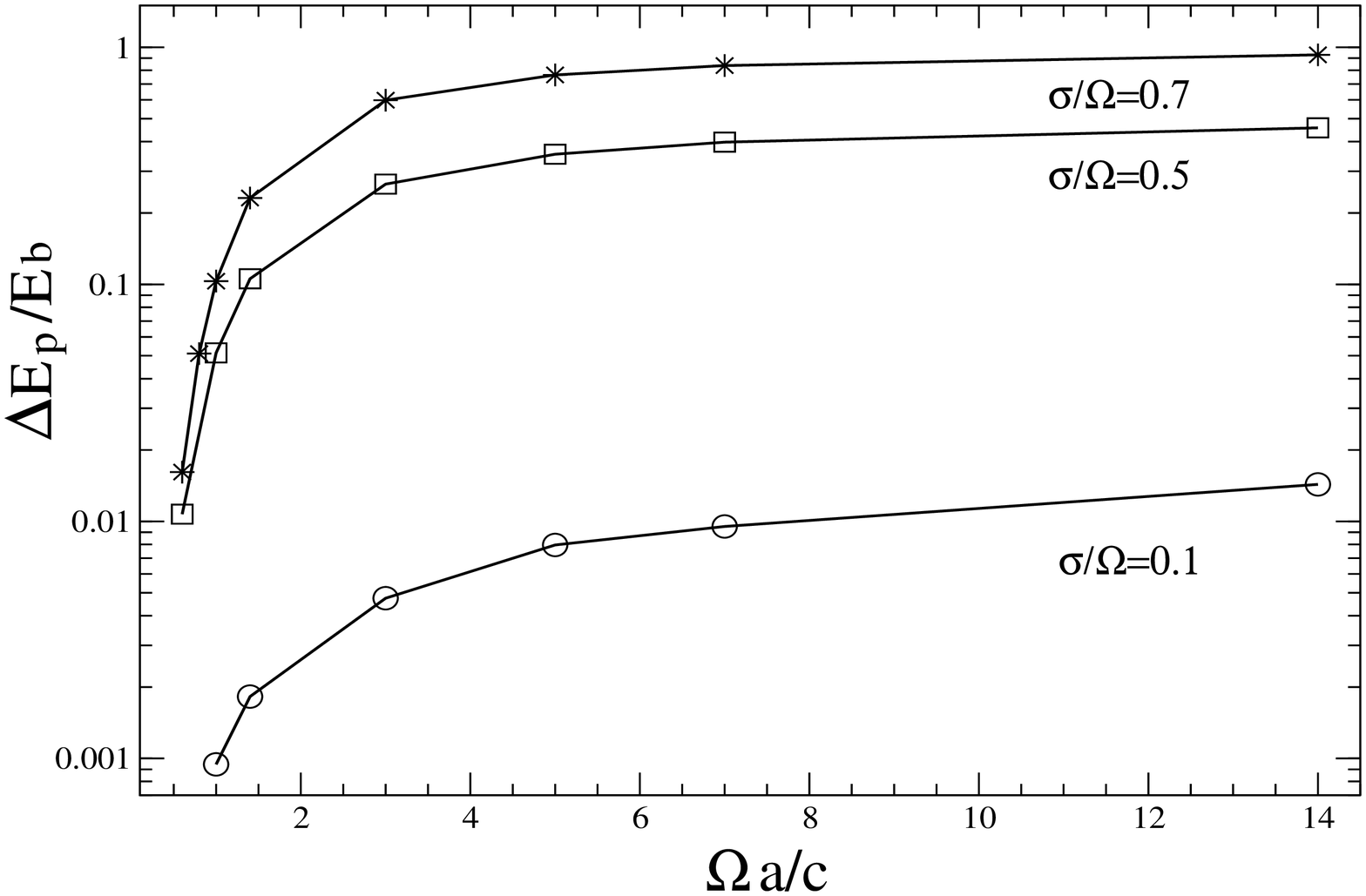}
\caption{Efficiency of superradiance, defined as the ratio between
  $\Delta E_{\rm p}=E_{\rm p}(\infty)-E_{\rm p}(0)$ and the background
  energy $E_{\rm b}$, plotted on a semi-logarithmic scale as a
  function of the angular frequency $\Omega$ (in units of $c/a$) for
  three values of the ratio $\sigma/\Omega$.}
\label{Fig8}
\end{figure}

Finally, the superradiant behavior of the acoustic black hole during the
scattering of the wavepacket is illustrated in Fig. \ref{Fig7} and
\ref{Fig8}. In Fig.\ref{Fig7} we show a typical time evolution of 
the energy of wavepacket 
$E_{\rm p}(t)=(\rho M/2)\int_0^{2\pi} \,{\rm d}\phi \int_{0}^{H}\, {\rm d}z
\int_{a}^{143a} v_1^2 r\,{\rm d}r$ with $v_1=\nabla \theta^{(1)}$,
$\rho$ the density and $H$ the axial extent  of the fluid, normalized
to its initial value $E_p(0)$, for $\sigma=0.7c/a$ and 
$\Omega=1.4c/a$, within the superradiant regime ($m=1$, solid line) and 
outside it ($m=0$, dashed line). 
In the non-superradiant case, the energy of the scattered 
wavepacket goes asymptotically to zero, indicating 
that all the impinging energy is lost  
to the vortex sink.   
In the superradiant case instead,
the energy of the back-scattered wavepacket exceeds its initial value, 
showing  extraction of  energy from the  
ergosphere at the expense of the rotational kinetic energy. 
The energy gained via 
superradiance is seen to 
exceed in this case twenty percent of the initial value $E_p(0)$. 

In Fig. \ref{Fig8} we show the superradiance efficiency, defined as
 the ratio between the total energy gain
$\Delta E_{\rm p}=(E_{\rm p}(\infty)-E_{\rm p}(0))$ of the wavepacket
 and the background energy $E_{\rm b}$, for three
values of $\sigma/\Omega$ in the superradiant range, 
as a function of $\Omega a/c$.
In the perturbative regime (for $\Omega\leq 3c/a$, say) the
efficiency of energy extraction grows very rapidly 
with $\Omega$, especially at large values of the ratio
$\sigma/\Omega$ between the central frequency of the
 quasi-monochromatic wavepacket and
 the angular rotation frequency of the vortex.
Although substantial values of the efficiency are beyond the scope of 
the perturbative Klein-Gordon  
description used in this work, the indication from Fig. \ref{Fig8} is
 that an efficiency approaching unity may be obtained by a suitable
 match between the frequency of the wavepacket and the rotational
 frequency. Indeed, it seems that nonlinear
effects may primarily determine the way in which the energy extraction
behaves as it becomes comparable to the background energy.

\section{Discussion}

Interest in pursuing analogue models of gravitational physics has been
steadily growing in the recent past. Our attention in the present work
has been directed to sound wave propagation in a background fluid flow
as an analog for field propagation in a curved space-time, as first
highlighted in the seminal work of Unruh \cite{ref41}. As a first step
in a long-term program, we have studied the scattering and radiance
phenomena from a black hole whose background space-time may be
modelled by the flow-field configuration associated with a vortex
excitation in a fluid with a drain. We have deliberately chosen values of the system
parameters that are typical of superfluids in a Bose-Einstein
condensed state. The inverse crossing time of a vortex in such a
dilute gaseous superfluid is $c/a\approx 15 KHz$, of the same order as
for a cosmic black hole of radius $a\approx 10 km$. Such a close match
is a consequence of the very low speed of sound in condensates, which
is a few mm/s.

Our analysis of the acoustic analog has shown that acoustic black
holes behave under several respects like black holes in GR. They have
an event horizon and an ergosphere, and can superradiate at the
expense of their rotational kinetic energy. Once excited, they come
back to their initial state and are stable against sharp
perturbations. The special methods that one typically uses in the
theoretical treatment of black holes in space-time can be applied here
 in a straightforward way. Our calculations have indicated that substantial
superradiance efficiencies may be achieved from a 
\lq\lq terrestrial black hole\rq\rq  by a proper choice of the system parameters. This suggestion will have to be substantiated by the inclusion of nonlinear and quantum effects, and indeed the acoustic analogy is still awaiting for
a deeper analysis and understanding.

An especially important direction for further applications of
numerical relativity to superfluids will be to develop a
proper inclusion of vortex quantization. As is well known, the
vorticity of a superfluid is quantized in units of the
circulation $\kappa=h/M$, {\it i.e.} $\Omega_n=nh/(Ma^2)$, where $n$
is an integer, $M$ the particle mass, and $a$ the size 
of the vortex core. Consistency requires that an integer number of quanta be exchanged between a vortex and a wavepacket in a scattering process. A semiclassical treatment is thus justified only in the case of scattering from so-called giant vortices built from up to $n\simeq 60$ quanta, as are indeed observed in some experiments on Bose-Einstein condensates inside strongly anharmonic confinements (see {\it e.g.} \cite{giant}). It will also be necessary in this context to consider a vortex with no drain, such a configuration being accessible in the linear regime by an immediate modification of the present acoustic analog \cite{SAVAGE,Volovik}. More generally, the inclusion of both nonlinear and quantum effects in wavepacket propagation inside superfluid flows of dilute quantum gases can be based on the use of the time-dependent Gross-Pitaevskii equation. Advanced techniques are available for its numerical solution in a variety of configurations \cite{Minguzzi}. Work along these lines is in progress.

\begin{acknowledgments}
This work was partially supported by an Advanced Research Initiative of SNS.
\end{acknowledgments}

\appendix \section{Properties of the acoustic metric in equation (3)\label{appA}}

The line element (\ref{METRIC1}) is most conveniently
analysed by recasting it in the following $(3+1)$-form:
\begin{equation}
{\rm d}s^2=-\alpha^2{\rm d}t^2+\gamma_{ij}({\rm d}x^i+\beta^i {\rm
  d}t)({\rm d}x^j+\beta^j {\rm d}t)\,.
\end{equation}
The three-metric is the flat Euclidean space
 in cylindrical coordinates $\gamma_{ij}={\rm Diag}(1,r^2,1)$.
Latin letters are used for spatial indices that are raised and lowered with $\gamma_{ij}$ and its inverse $\gamma^{ij}={\rm Diag}(1,r^{-2},1)$, whereas 
Greek letters indicate four-dimensional quantities.
The shift is given by $\beta^i=({ a c}{r^{-1}}, -{ a^2\Omega}{r^{-2}},
 0)$, and consequently $\beta_i=({ a c}{r^{-1}}, - a^2\Omega, 0)$. 
The lapse is $\alpha= c$.

The slices $t={\rm const}$ are flat, and consequently this effective space-time is written in Painlev\'e$-$Gullstrand-like form. The presence of a shift vector manifests the well know phenomenon of frame dragging.
The metric depends on the radius only, so that three Killing vectors
are associated to cyclic coordinates.
This, together with Group Theory, ensures that both geodesics (sound rays) and wave fields can be solved by separation of variables.
Moreover this Killing vector is not rotation free, {\it i.e.} the space-time is stationary.
The metric, which diverges  in the origin only, possesses an event 
horizon, and in order to demonstrate this assertion 
we have to analyse the null geodesics.

We start by writing the principal function
$S(t,r,\phi,z)=-Et+f(r)+L\phi+Qz$, where the quantities $E$,$L$, and $Q$
are constants of the motion.
The associated Hamilton-Jacobi equation $\frac12\nabla^\mu S\nabla_\mu S=0$ for massless particles becomes
\begin{eqnarray}
&&\left(1-\frac{a ^2}{r^2}\right)\left(\frac{{\rm d}f}{{\rm
    d}r}\right)^2-\frac{2a \left(Er^2-a ^2\Omega
  L\right)}{cr^3}\frac{{\rm d}f}{{\rm
    d}r}+\\
&&\left[Q^2+\frac{L^2}{r^2}-\frac{E^2r^4-2a ^2\Omega L
    Er^2+L^2\Omega^2a ^4}{c^2r^4}\right]=0\nonumber
\label{elq}
\end{eqnarray}
and is solvable by quadrature in terms of elliptic integrals.
The gradient of $S$ leads to the quantity $U^\mu=\nabla^\mu S$, which
is conventionally called the four-velocity of the null
geodesics, although for light-rays it should more
appropriately be called the four dimensional wave vector $k^\mu$.

If we assume no motion along the $z$ axis, {\it i.e. }  $Q=0$, and zero
angular momentum $L=0$ (at infinity the motion is purely radial),
Eq. (\ref{elq}) above delivers the two solutions
\begin{equation}
f_{1,2}(r)=\mp\frac{E}{c}r+\frac{Ea }{c}\log\left({\frac{r}{a }\pm1}\right).
\end{equation}
The associated four velocities are
\begin{eqnarray}
U_1^\mu\equiv n^\mu &=&\left[\frac{rE}{c^2(r+a
    )},-\frac{E}{c},\frac{E\Omega a ^2}{rc^2(r+a )},0\right],\\
    U_2^\mu\equiv l^\mu &=&\left[\frac{rE}{c^2(r-a
    )},\frac{E}{c},\frac{E\Omega a ^2}{rc^2(r-a )},0\right]\nonumber.
\label{PNDS}
\end{eqnarray}
The quantity $n^\mu$  is the  vector tangent to an ingoing congruence
of null geodesics, and the quantity $l^\mu$ corresponds instead to the outgoing ones.
The ingoing null signals are regular on the surface $r=a $, whereas
the outgoing ones diverge. This means that at $r=a $ every signal
can enter but not escape, {\it i.e}. there is an event horizon.
This pair of congruences of null vectors are the principal null
directions associated with this geometry. More in detail, they
automatically  bring the Weyl tensor into its canonical form, {\it i.e.} a
Petrov type-D manifold as for stationary black
holes in General Relativity. However, unlike GR cases, the
Ricci scalar is non-zero, implying that the Goldberg-Sachs theorem
for type-D space-times does not apply.

A new coordinate system for this acoustic geometry can be tailored to
the ingoing light rays associated with these vector fields. This step
is easily performed by defining a new set of coordinates $v$ and
$\psi$, which are related to the old ones by
\begin{eqnarray}
t&=&\frac{v}{c}-\frac{r}{c}+\frac{a }{c}\log\left(1+\frac{r}{a
}\right),\\
\phi&=&\psi-\frac{\Omega a }{c}\log\left(\frac{r}{r+a
}\right)\nonumber.
\label{COORDYS}
\end{eqnarray}
Geometrically, the new coordinates compress time and untwist 
the angle in the neighborhood of the horizon.
The corresponding metric simplifies to
\begin{eqnarray}
{\rm d}s^2=&-&\left(1-\frac{a ^2}{r^2}-\frac{a ^4\Omega^2}{c^2
  r^2}\right){\rm d}v^2+2{\rm d}v{\rm d}r\\
&-&\frac{2\Omega a ^2}{c}{\rm
  d}v{\rm d}\psi+r^2{\rm d}\psi^2+{\rm d}z^2\nonumber, 
\label{BELLAMETR}
\end{eqnarray}
and the ingoing congruence becomes  $n^{\mu}=(0,-E,0,0)$, as
 expected. 
 The space-time is now expressed in \lq\lq Kerr-ingoing like\rq\rq coordinates.

The vanishing of the $g_{tt}$ component of the metric tensor means
that the norm of the temporal Killing vector
$\xi^\mu=\left(\frac{\partial}{\partial t}\right)^\mu$ changes sign at
the radius $r_{erg}=\frac{a }{c}\sqrt{c^2+a ^2\Omega^2}$. This manifests the presence of an ergosphere and suggests the possibility of rotational energy extraction via superradiance.

\end{document}